\documentclass[aps,prx,twocolumn,showpacs,superscriptaddress,floatfix]{revtex4-2}

\usepackage[titletoc,toc,title]{appendix}
\usepackage{graphicx}
\usepackage{bm}
\usepackage{xcolor}
\usepackage{booktabs}
\usepackage{hyperref}
\usepackage{diagbox}
\usepackage{braket}
\usepackage{amsmath}
\usepackage{chngcntr}
\renewcommand{\vec}[1]{\bm{#1}}
\begin{document}

\title{Seeing moir\'e: convolutional network learning applied to twistronics}

\author{Diyi Liu}
\affiliation{School of Mathematics, University of Minnesota—Twin Cities, Minneapolis, Minnesota 55455, USA}
\author{Mitchell Luskin}
\affiliation{School of Mathematics, University of Minnesota—Twin Cities, Minneapolis, Minnesota 55455, USA}
\author{Stephen Carr}
\affiliation{Department of Physics, Brown University, Providence, Rhode Island 02912-1843, USA}
\affiliation{Brown Theoretical Physics Center, Brown University, Providence, Rhode Island 02912-1843, USA.}

\date{\today}

\begin{abstract}
Moir\'e patterns made of two-dimensional (2D) materials represent highly tunable electronic Hamiltonians, allowing a wide range of quantum phases to emerge in a single material.
Current modeling techniques for moiré electrons requires significant technical work specific to each material, impeding large-scale searches for useful moir\'e materials.
In order to address this difficulty, we have developed a material-agnostic machine learning approach and test it here on prototypical one-dimensional (1D) moir\'e tight-binding models.
We utilize the stacking dependence of the local density of states (SD-LDOS) to convert information about electronic bandstructure into physically relevant images.
We then train a neural network that successfully predicts moir\'e electronic structure from the easily computed SD-LDOS of aligned bilayers.
This network can satisfactorily predict moir\'e electronic structures, even for materials that are not included in its training data.
\end{abstract}

\maketitle

\section{Introduction}


Since the discovery of strongly correlated electronic phases in twisted bilayer graphene (TBG)~\cite{Cao2018mott,Cao2018sc}, moir\'e materials made of multiple layers \cite{Zhang2021,Park2021,Hao2021,Kim2022,Shen2020,Rickhaus2021} or 2D crystals other than graphene~\cite{Sun2021,Xu2020,Wang2020,Ghiotto2021} have been under intense study.
A primary goal in these efforts is to understand how the rich phase diagram of TBG can be realized in other materials, and if phases distinct from those which appear in graphene superlattices are possible.
However, the theoretical modeling or experimental realization of all possible 2D bilayer materials is not an easy task.
Even if only the materials in which only a few atoms per unit cell are considered, there are thousands of structures with exfoliating layered geometries~\cite{mounet2018two}.
First principles calculations of moir\'e superlattices are extremely computational demanding~\cite{Lucignano2019}, and existing effective models are only designed for simple one band or two band bilayer systems~\cite{Carr2020review,Bistritzer2011,Wu2018,Wu2019,Hao2021TI}. 
Therefore, high-throughput screening is a necessary approach to help the community focus on materials that most likely possess the deserved properties.
Computing electronic properties from first principles on only a small number of promising bilayer materials is much more practical than blindly performing such expensive calculations on every possible bilayer material. 

\begin{figure}
    \centering
    \includegraphics[width=\linewidth]{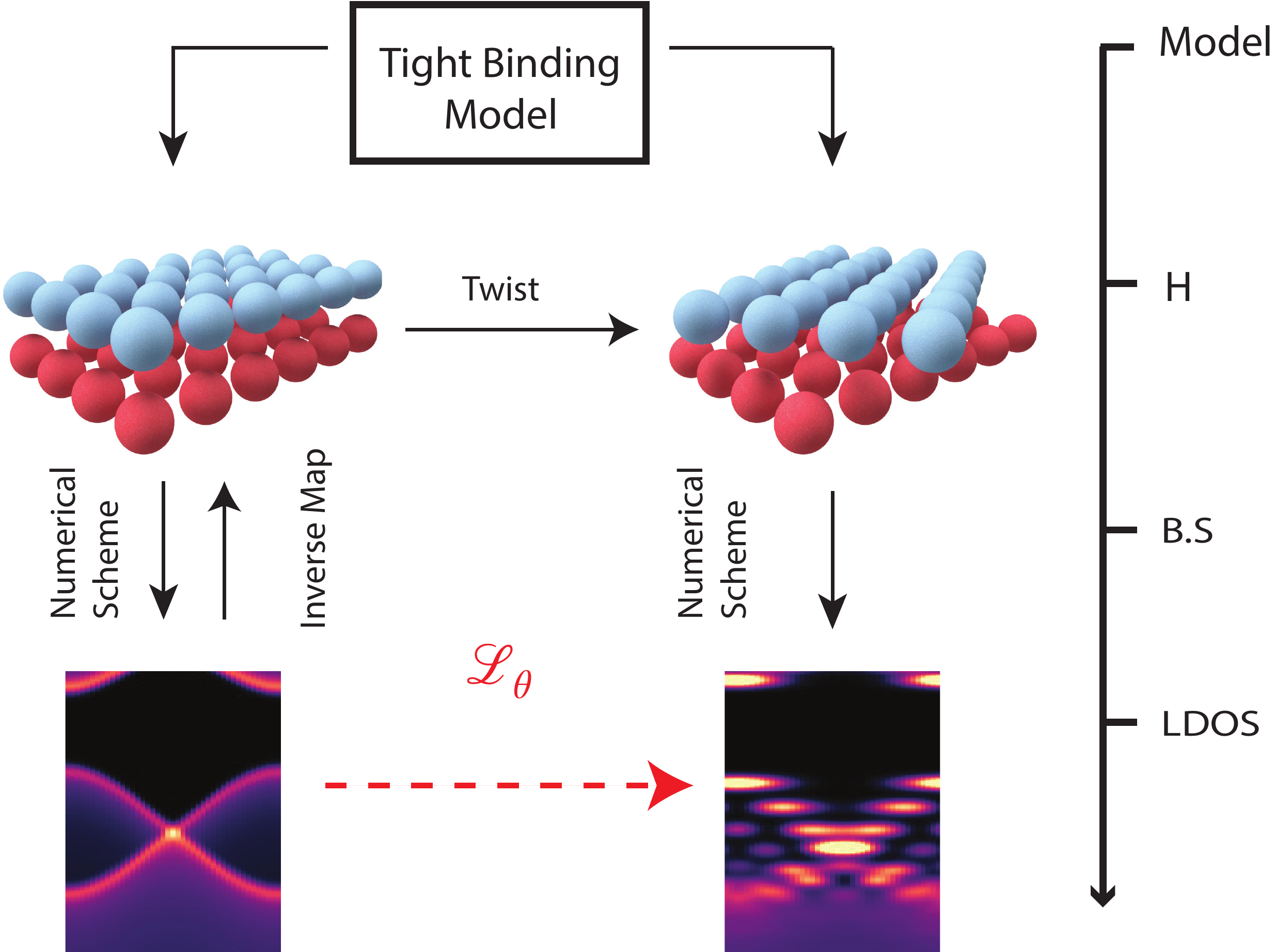}
    \caption{A non-commuting diagram of electronic structure models, showing the connections between the (top) parameterization of a tight binding model for a 2D material, the (middle) explicit tight-binding Hamiltonians (H) and resulting band structure (B.S.) at specific shifts or twist angles, and (bottom) the configuration-dependent electronic local density of states (LDOS). The $\mathcal{L}_{\theta}$ marked with the red dashed arrow is the image processing we aim to achieve, e.g., the Twist Operator to be learned with a neural network. The solid black arrows represent well-defined mappings.}
    \label{fig:diagram}
\end{figure}

Putting first principle calculations aside, even tight binding calculations of twisted bilayer systems are not straightforward. 
Accurate models for the individual layers and for their interlayer coupling must first be obtained, and at small angles large matrices ($>50,000$ electrons) occur for systems with many $d$-orbital electrons, making diagonalization difficult.
Although fast numerical schemes based on plane-wave approximations of large moir\'e superlattices have been developed for the computation of electronic properties of twisted bilayers \cite{massatt2017electronic,massatt2017incommensurate,Wu2018,Hao2021TI}, these still rely on accurate tight-binding models which can be challenging to generate for arbitrary materials.
However, what if the changes of electronic structures from an untwisted to a twisted heterostructure followed certain universal rules, dependent only on the twist angle $\theta$?
Assuming such a transformation exists, it can be characterized by a material-agnostic ``Twist Operator,'' and one could overcome both the above challenges.
Tight-binding parameterization would no longer be necessary, as one can directly pass untwisted electronic structure to the operator.
Large computational complexity is avoided in two ways.
First, one only needs to perform calculations on aligned unit-cells at various configurations, which can be performed easily and quickly with first principle methods.
Second, once the training necessary for learning the operator is complete, the resulting model will by many orders of magnitude faster than a direct tight-binding or DFT approach to the moir\'e electronic structure problem.
We present such a method here, and show that we can accurately approximate a universal twist operator with a convolutional neural network, with the overall concept outlined in Fig. \ref{fig:diagram}.
For this first attempt at learning the twist operator, we focus on a making predictions at a single ``twist angle'' (lattice mismatch in 1D) of $\theta = 0.1$. 
This network is able to make predictions for bilayer materials absent from its training set, but not for materials too different any represented in the training. We posit that this method can become a powerful tool for high-throughput screening of proposed structures or even directly applied to experimental data.

Learning a generalized operator with neural networks remains a hard problem even in recent years.
The main challenges of learning an operator depend on the nature of the input space and the output space, both of which are infinite-dimensional for highly-nonlinear operators.
However, universal approximation theory gives some hope that certain operators can be learned by neural networks, namely any nonlinear \textit{continuous} operator~\cite{chen1995universal,lu2019deeponet}.
Most recent applications of operator learning focus on solving a nonlinear partial differential equation \cite{bhatnagar2019prediction,li2020fourier,li2020neural} and reconstruction of a coefficient function in the differential operators. 

\begin{figure}
    \centering
    \includegraphics[width=\linewidth]{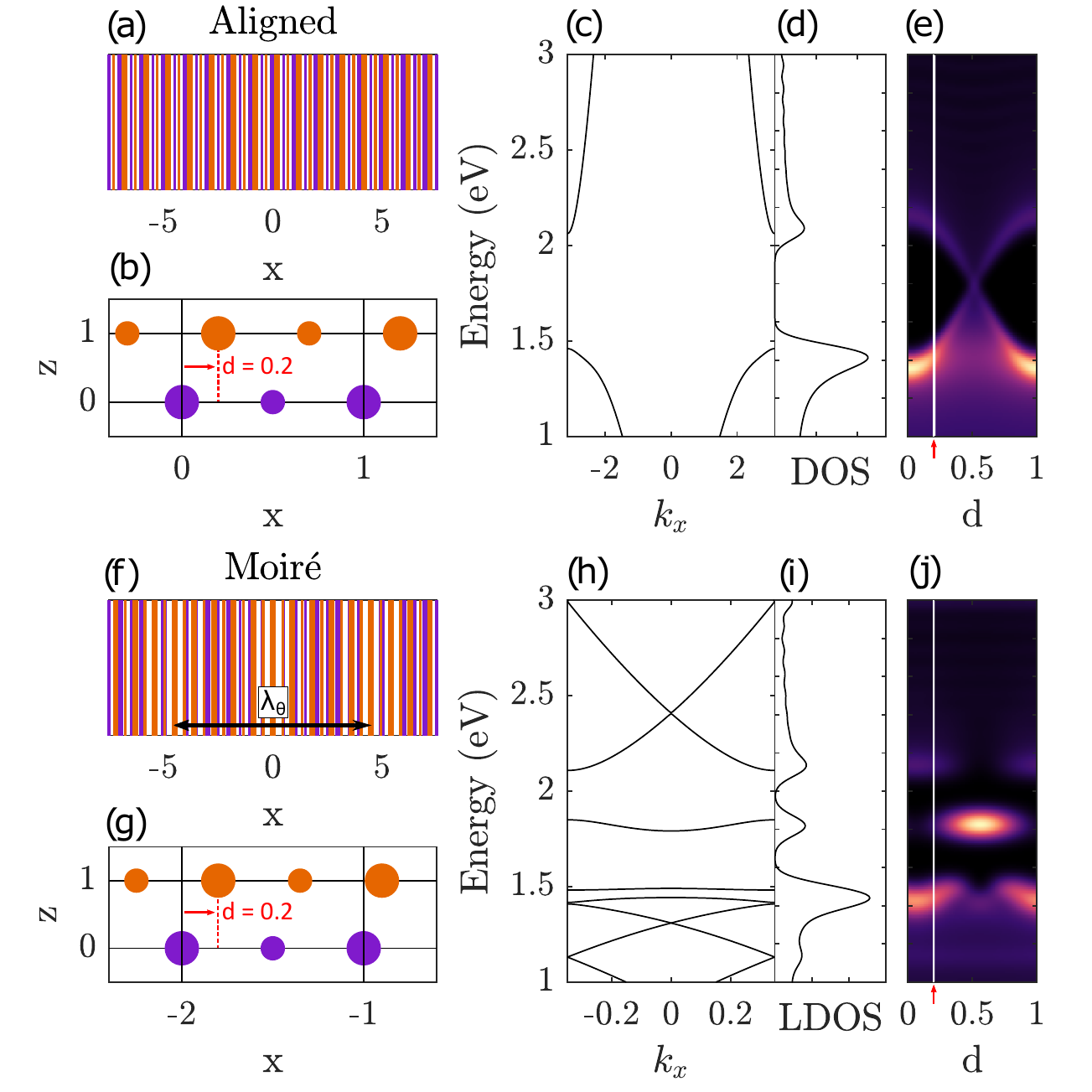}
    \caption{a) Top-down view of an aligned 1D bilayer, with atomic sites represented by vertical lines. The top layer is in orange while the bottom layer is in purple b) A zoomed-in side view showing the atomic sites as circles at the specific configuration ($d$) between the top and bottom layers c) band structure and d) LDOS for this aligned bilayer system and a moir\'e bilayer system. e) Configuration-dependent DOS map, for all possible shifts between the two layers in the aligned geometry, with the selected shift in a) highlighted with the white vertical line.
    f-j) same as (a-e), but for a 1D moir\'e pattern with $\theta = 0.1$ and the 1D moir\'e length indicated in (f) by $\lambda_\theta$.
    Note that in b), the two layers are uniformly shifted by $d=  0.2$, but in g) the two layers are aligned at $x = 0$ and then the top layer is compressed.}
    \label{fig:intro}
\end{figure}

In this work, we frame the learning of the twist operator in a manner analogous to an image processing problem.
The electronic structure of the untwisted and twisted bilayers are transformed into ``images'' by plotting their stacking-dependent local electronic density of states (SD-LDOS)~\cite{Carr2017,carr2020duality}, in contrast to the conventional approach of plotting the momentum variation of eigenvalues (band structure). 
This so-called configuration space acts as a compact domain for understanding moir\'e electronic structure.
In particular, during the conversion of band structures all moir\'e flat-bands become ``bright spots'' in the resulting images (Fig. \ref{fig:intro}).
To simplify this first attempt at generation of a twist operator, we limit ourselves to a collection of ten artificial 1D materials, and mimic the role of the twist angle in 2D materials with a 1D lattice mismatch to generate moir\'e patterns (Fig. \ref{fig:bands}).
In Section \ref{sec:methods}, we introduce the electronic structure of these moir\'e 1D materials, the generation of the configuration-space LDOS maps, and our neural network.
In Section \ref{sec:results}, we cover the results of various neural network trainings, including optimization of the network and material-agnostic prediction of moir\'e electronic structure from ``untwisted'' reference data. Finally, we summarize these results and discuss future applications of this methodology in Section \ref{sec:conclusion}.

\section{Methods}
\label{sec:methods}

\subsection{1D moir\'e electrons}
To test the viability of using a neural network for predicting moir\'e electronic structure, we will consider 1D bilayer chains.
Although most 2D moir\'e materials are generated by introducing a relative twist between the two layers of a bilayer, it is not possible to ``twist'' two 1D chains.
Instead, we use a relative compression or expansion of the lattice parameter, which can also generate a moir\'e superlattice, following recent studies of 1D moir\'e patterns \cite{carr2020duality,Tritsaris2021}.
We define this lattice mismatch by a parameter $\theta$ (\textit{not} a twist angle, but with this variable chosen to facilitate conceptual comparison to twisted 2D systems), such that the bottom layer has lattice-parameter $1$ and the top layer has lattice parameter $1-\theta$.
We have chosen length to be unitless for notational simplicity.
In Fig. \ref{fig:intro}, we compare the geometry and electronic structure between a one dimensional bilayer with and without a lattice parameter mismatch.
Throughout this work, we fix $\theta = 0.1$, corresponding to a periodic moir\'e supercell of 10 and 11 unit-cells for the bottom and top layers, respectively.
For both $\theta = 0$ and $\theta = 0.1$, we calculate the electronic band structure.
Just as in the case of twisted 2D bilayers, the 1D moir\'e pattern introduces multiple moir\'e minibands, with the flattest bands near the band gap.

Similar information can be gained by considering the stacking dependent local density of states (SD-LDOS).
To understand how the electronic structure varies under different stacking configurations, we first consider a bilayer system (either aligned or mismatched by $\theta$) such that the unit cell corner of the top and bottom layer are vertically aligned at the origin.
We label the top layer's unit-cell nearest to the origin $R_0$, and then allow this initially aligned unit-cell to move left or right by an amount $d$.
This produces a bilayer system with the desired stacking $d$ near the origin (Fig. \ref{fig:intro}a,b), and to obtain the SD-LDOS we need only consider the orbitals in $R_0$.
The SD-LDOS is then defined as:
\begin{equation}
\rho_{\theta,d}(E) = \frac{1}{N_k} \sum_{o \in R_0} \sum_{n,k} |\psi^{nk}_o|^2 \delta(E - E_{nk})
\label{eq:LDOS}
\end{equation}
where $o$ are the orbitals in the top-layer unit-cell $R_0$ with stacking $d$, $n$ indexes the bands of the system, $k$ indexes the momentum states of the 1D Brillouin Zone, and $N_k$ are the number of momentum points sampled. The eigenpairs $\{\psi^{nk}_o$,$E_{nk}\}$ are obtained from a Hamiltonian $H_k(\theta,d)$, which is a moir\'e superlattice with lattice mismatch $\theta$ and with the top layer shifted $d$ relative to the bottom layer at $R_0$.
Note that for the aligned system, the sum over $o \in R_0$ is equivalent to summing over all orbitals in the top-layer, as the system is periodic within a single lattice length.
Comparing the $\theta = 0$ to the $\theta \neq 0$ case in Fig. \ref{fig:intro}c,\,d, we see that the energy regions that had large variation with $d$ in the aligned system change the most after the introduction of a moir\'e pattern.
Importantly, the moir\'e pattern has converted an ``X'' shape in the aligned system's SD-LDOS into a state that is localized in space and energy ($d = 0.5$ and $E = 1.75$ eV).

\begin{figure}
    \centering
    \includegraphics[width=\linewidth]{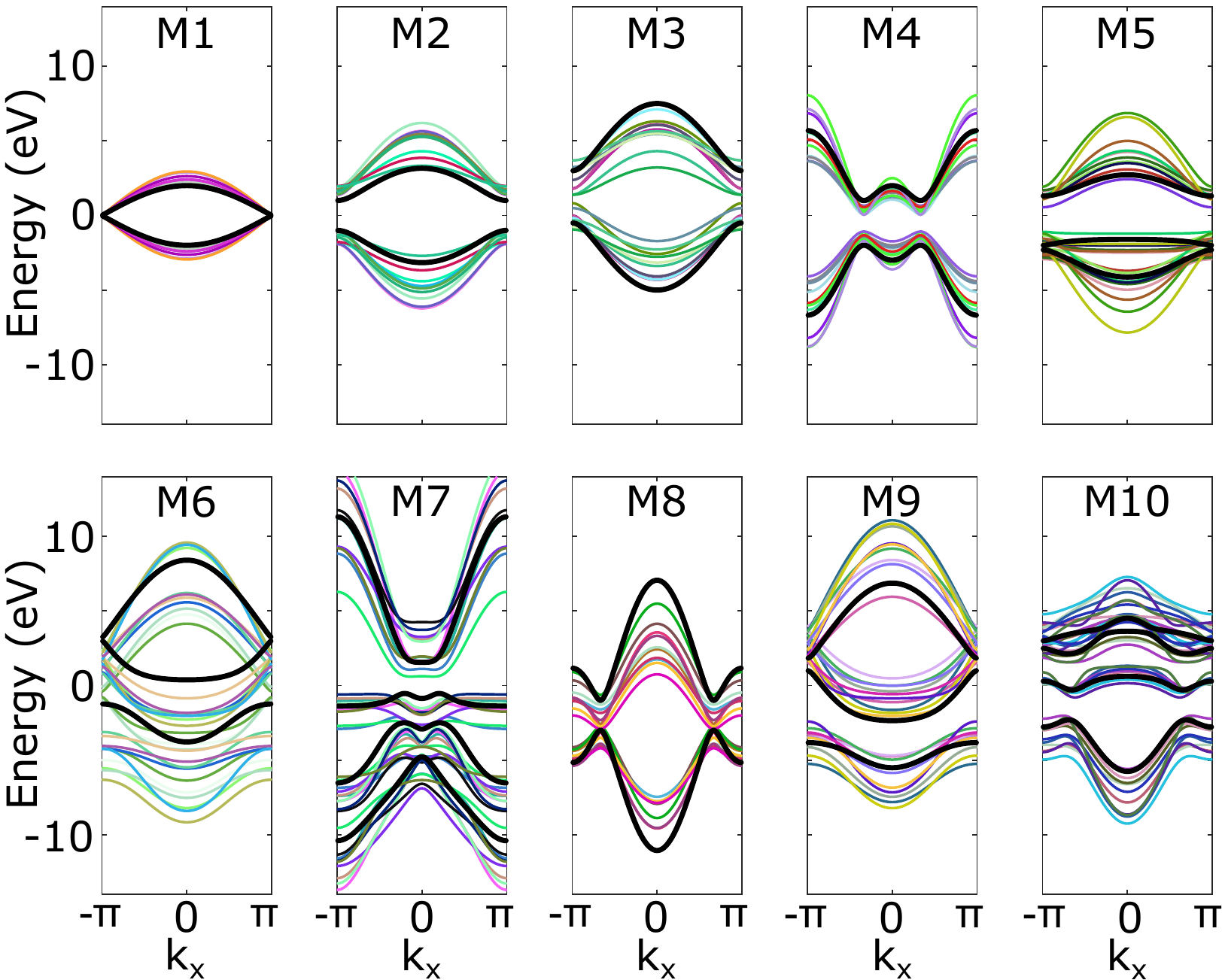}
    \caption{One-dimensional band structures for the ten Hamiltonian templates, M1-M10, with randomized parameters. In black is a default setting for each Hamiltonian, and in color are examples of the same Hamiltonian with randomized tight-binding parameters.}
    \label{fig:bands}
\end{figure}

We chose ten prototypical classes of tight-binding models for our study, with their monolayer band structures shown in Fig. \ref{fig:bands}.
Each model represents a different type of material symmetry, and we  label them M1 through M10.
The first material, M1, is a two-orbital semi-metal, inspired by graphene. The next three, M2-M4, and M8 are two-orbital materials with a broken sublattice symmetry, inspired by hexagonal Boron Nitride (hBN).
The remaining materials are multi-orbital systems with a gap and complicated band hybridizations, inspired by transition metal dichalcognides, with M7 and M10 being the most complicated while M5,\,M6, and M9 are simpler three-orbital semiconductors.
We also introduce parameters of the interlayer tunneling function, which is defined as:

\begin{equation}
\label{eq:interlayer_t}
    t(\vec{r}) = \nu e^{-(r/R)^2}
\end{equation}

for $\vec{r}$ the distance between the two orbitals, $\nu$ the interlayer tunneling strength and $R$ the interlayer tunneling length.

Although ten materials would be quite a large number in the context of the study of moir\'e systems, for machine learning it is not diverse enough to allow for proper training of our model.
Therefore, we randomize the tight-binding and interlayer tunneling parameters around a set of base values.
Each material type consists of a Hamiltonian structure and set of random monolayer parameter distributions, which are sampled from for each individual element of the datasets.
That is to say, two samples of the same material type are never identical.
For the interlayer tunneling parameters, we also set $\nu = 0.3 \pm 0.1$ eV and $R = 0.5 \pm 0.1$, with uniform random sampling for each element of the dataset.
A full description of the monolayer Hamiltonians and their randomization is provided in App. \ref{app:hams}, and the code which generates the Hamiltonians and performs the neural network training is made publicly available~\cite{git_matlibrary}.

\subsection{The twist operator}
We now assert that the transformation from the aligned electronic structure to the moir\'e electronic structure is performed by a generalized twist operator, $\mathcal{L}_{\theta}$.
This operation is usually performed by applying conventional electronic structure methods to a moir\'e superlattice, be it with full DFT, a tight-binding model, or a plane-wave expansion of the monolayer Bloch states~\cite{Carr2020review}.
In Fig. \ref{fig:diagram}, starting with an accurate tight-binding parameterization of a 2D material, one can generate an aligned or twisted tight-binding model ($H$) and then calculate their respective band structure (B.S.).
In this context, the twist operator is not an explicit mapping which takes untwisted band structure into twisted band structure, but rather represents the underlying tight-binding parmeterization shared by both systems and the methodology of generating a twisted tight-binding supercell.
For another example of this scheme, consider the generation of a continuum model from stacking-dependent DFT calculations~\cite{Jung2014}.
In this case, the twist operator is a more explicit transformation from aligned information to moir\'e structure, but it is highly material-specific: the scientist must properly identify and extract the monolayer masses (variation in band energy with $k$) and moir\'e potentials (variation in band-gap with $d$) correctly, and then calculate the band structure in a plane-wave basis.
The implementation of this methodology to a new class of materials or symmetry point is quite the undertaking, as evidenced by the fact that each such extension is worthy of publication in itself \cite{Wu2018,Hao2021TI,Angeli2021}.

Our aim here is to approximate the twist operator by a convolutional neural network (CNN) applied to the SD-LDOS images, as marked with a red dashed arrow in the Figure \ref{fig:diagram}.
In support of $\mathcal{L}_{\theta}$'s existence, note that the effect of twisting a bilayer system is well defined and deterministic given the existence of a definite electronic structure model and a chosen twist angle.
Also note that the transformation, when viewed in the context of its operation on a real space Hamiltonian, is analytical with respect to $\theta$: the moir\'e pattern causes smooth variation in the local stacking and all current studies of moir\'e interfaces have found that the local Hamiltonians are likewise smooth in the stacking order~\cite{Jung2014,Wu2018,Wu2019,Hao2021TI}.
But to thoroughly prove the twist operator's existence, one must explore how electronic structure information is embedded in the SD-LDOS.
If we assume an inverse map from the SD-LDOS to the tight-binding model which generated it exists, then we can transform the SD-LDOS of an untwisted system to that of the twisted system by the conventional moir\'e supercell tight-binding method.
The geometric operation of twist and the calculation of LDOS from band structure are material agnostic, so all the material-dependent information is encoded in the inverse mapping of SD-LDOS to the tight-binding parameters.

Although the space of possible tight-binding models is infinitely dimensional, as long as numerous aligned LDOS values are known it is not unreasonable to expect some recovery of the parameters of tight-binding model are possible.
We note that this reconstruction of a tight-binding model from LDOS is \textit{not unique}, evidenced by the fact that there is a gauge choice present in the Wannierzation process \cite{Marzari2021}.
However, the gauge choice does not affect any physical observables, and so one could hope that the gauge choice does not effect the efficacy of mapping LDOS to SD-LDOS.
Detailed understanding of the necessary assumptions and resulting errors in this inverse mapping to a tight-binding model will be useful in optimizing our model, but as this work already presents a successful implementation we will leave a rigorous mathematical description of $\mathcal{L}_{\theta}$ existence to future work.

\subsection{Dataset and CNN}

We calculate the SD-LDOS on a uniform grid in both energy ($E_i$) and configuration ($d_j$), leading to an 2D array of LDOS values $\rho_{ij}$.
We first calculate the band structure of each system on a uniform sampling of $N_k$ $k$-points, and then implement Eq. \ref{eq:LDOS} by replacing the $\delta$ function with a Gaussian broadening of $\sigma$ to each eigenvalue:
\begin{equation}
\rho_{ij} \equiv \rho_{\theta,d_j}(E_i) = \frac{1}{N_k} \sum_{o \in R_0} \sum_{n,k} |\psi^{nk}_o|^2 \frac{1}{\sigma \sqrt{2\pi}}e^{-\frac{(E_i - E_{nk})^2}{2\sigma^2}}
\label{eq:LDOSgaussian}
\end{equation}
with $\sigma = 30$~meV. The $\rho_{ij}$ can then be interpreted as images and the learning of the moir\'e SD-LDOS can naturally be seen as an image processing problem.

\begin{figure}
    \centering
    \includegraphics[width=\linewidth]{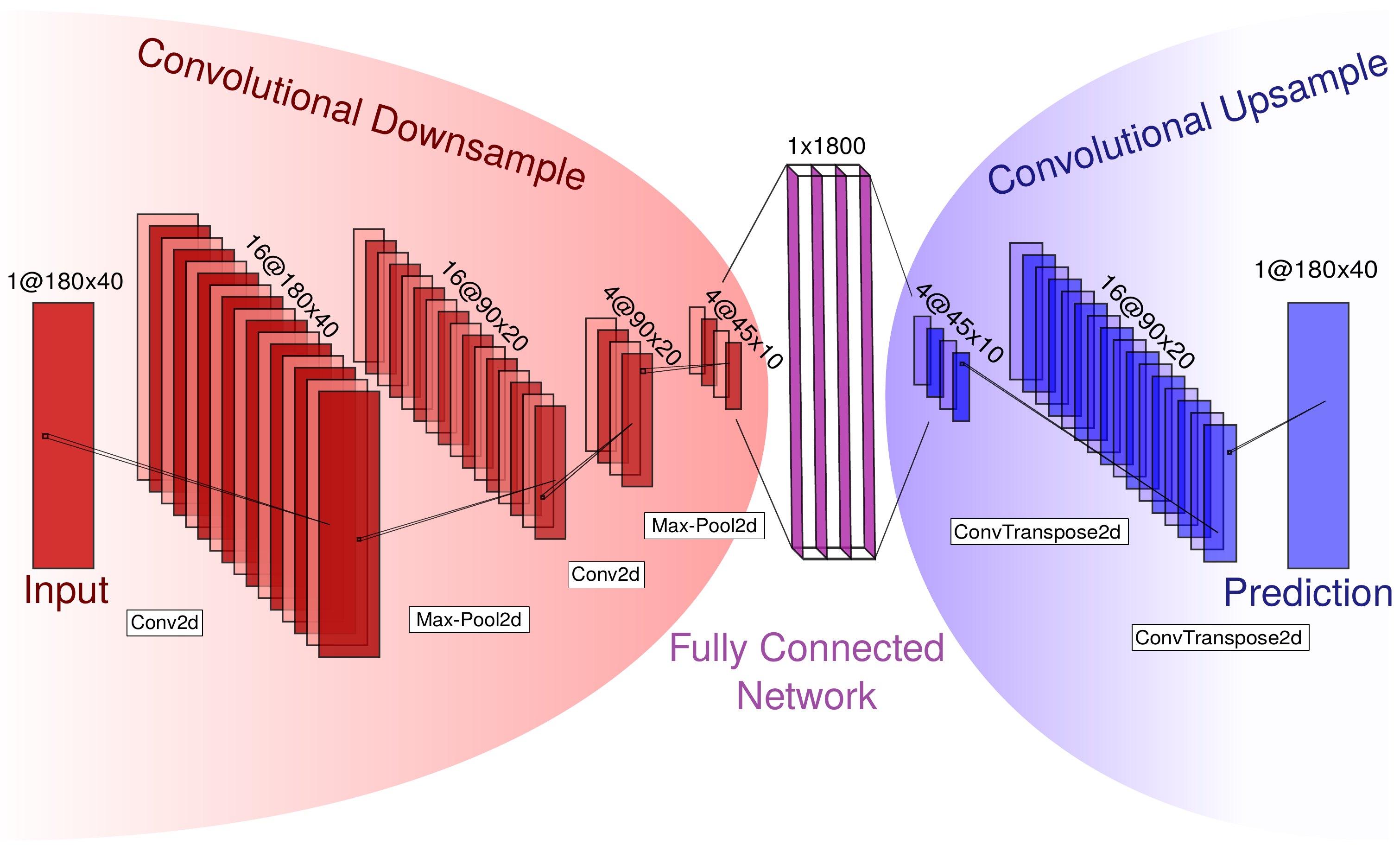}
    \caption{Selected network architecture, which consists of a fully connected neural network (for prediction) at the center of a convolutional autoencoder (for image compression/decompression). Each convolutional layer is given by $m@h \times w$ for $m$ nodes of pixel size $h \times w$, with the algorithm between each layer given below in white boxes. The number of layers $n$ in the fully connected network at the center is optimized to find an optimal complexity for our training duration, with $n = 4$ displayed here.}
    \label{fig:network_arch}
\end{figure}

In Figure \ref{fig:network_arch} we introduce our selected neural network architecture, which consists of three components including an encoder, a fully connected network, and a decoder.
This follows a convolutional auto-encoder structure, well known to provide effective solutions to image processing problems \cite{Ronneberger2015}.
We found that using such an encoder/decoder CNN structure decreases the training and test error by a factor of 10 compared to a fully connected neural networks of similar size, and requires much less training time.

In our network, the encoder plays the role of dimension reduction and is expected to compress the LDOS calculated from aligned bilayer system into some generalized parameters in the space of tight-binding models.
It takes as input a window of aligned LDOS, represented by a single channel $180\times 40$ image.
The encoder uses rectified linear unit (ReLU) activation function and it is composed of two convolutional layers and two pooling layers.
In the convolution layers, a $3\times 3$ kernel is used when both padding and stride are set to be one. Between the first pooling layer and second convolution layer, a batch normalization is applied.
A fully connected neural network is then used to work as the twisting operation, and represents the second component of the neural network architecture.
It consists of multiple dense layers and uses an ReLU activation function.
Batch normalization is applied between different fully connected layers.
The decoder structure, the last component of the network architecture, uses an upsampling CNN and ReLU activation function.
It is designed to resemble the process of calculating the LDOS from tight binding parameters for moir\'e bilayer systems.
Padding and stride in the decoder structure are set to be 0 and 2, respectively, in these layers, and batch normalization is applied between two upsample convolution layers.

For the success of any machine learning application, data preprocessing is often more important than the chosen architecture, and so we now discuss our choices of preprocessing thoroughly.
We normalize the SD-LDOS of each bilayer material by the maximum value of the aligned SD-LDOS.
This choice means that the moir\'e LDOS is allowed to be larger than $1$, allowing one to predict large DOS enhancement.
We next assume that the full spectrum of the LDOS is not necessary, and that one can learn local spectral structure instead of the global structure.
In other words, if the width of a truncated energy window is larger than any parameters in the underlying model, a local reconstruction of moir\'e structure from just the information in that window should be possible.
The SD-LDOS is calculated first for an energy range  $[-4.5, 4.5]$ eV and  different $20$ local configurations $d_j$ along the 1D unit cell. 
We then chose nine energy windows, each with a width of $1.8$ eV.
Those nine windows are generated by shifting the bottom energy window  up by $0.9$ eV eight times, such that a given energy never appears more than twice in our dataset for a specific material.

Our full dataset consists of ten types of bilayer materials.
To understand the transfer learning of the network, we also consider a dataset with M7 and M10 removed, and a dataset with M2, M5, M7, and M8 removed.
This first reduced dataset, labeled R1, has removed the two most complicated materials (M7 and M10), and tests how the network performs for materials that are vastly different than any member of its training set.
In contrast, the second reduced dataset, labeled R2, is designed to test how the network does on materials that are similar, but slightly different to other members of its training set.
Although the different datasets differ by their total number of materials, the training error and test error is normalized per batch. Therefore, the training and test error across datasets is still comparable.

The full dataset and two reduced datasets consists of 20000, 12000, and 16000 SD-LDOS images, respectively, with each included material type contributing equally.
During the training of our neural networks, 80\% of the data is used as training data with 20\% kept in reserve to serve as test (validation) data.
The optimization algorithm used in neural network training is Adaptive moment estimation (ADAM) \cite{kingma2014adam} with a fixed learning rate set to 0.001 and stochastic small-batch training scheme applied
For a loss function, we use the mean square error (MSE) on each truncated energy window of the SD-LDOS.
The MSE, or $L_2$ distance, between the true $\rho$ (obtained by tight-binding calculation of the moir\'e bilayer) and the CNN's predicted $\rho'$ is
\begin{equation}
\label{eq:error}
\Delta = \frac{1}{N}\sum_{ij} (\rho_{ij} - \rho'_{ij})^2.
\end{equation}
where $N = N_d N_E$ is the number of total points in the windowed LDOS. Here we also define the moir\'e sensitivity $\Delta_m$, which is the $L_2$ distance between the aligned and moir\'e LDOS.
$\Delta_m$ uses an identical formula as above but with $\rho$ and $\rho'$ corresponding to the input and output images respectively instead, with no reference to the network prediction.

To perform the training, each epoch consists of selecting an energy window at random and then applying ADAM across the entire training set in batches.
The training error for each epoch is calculated by summing the MSE of all batches.
The test error is calculated after each epoch finishes by summing all the MSE of equally sized batches over the whole testing set.
The two errors reflect how well the neural network model is learning the training data and how well the neural network predict results on the unseen data in the test set.

\section{Results}
\label{sec:results}
\begin{figure}
    \centering
    \includegraphics[width=\linewidth]{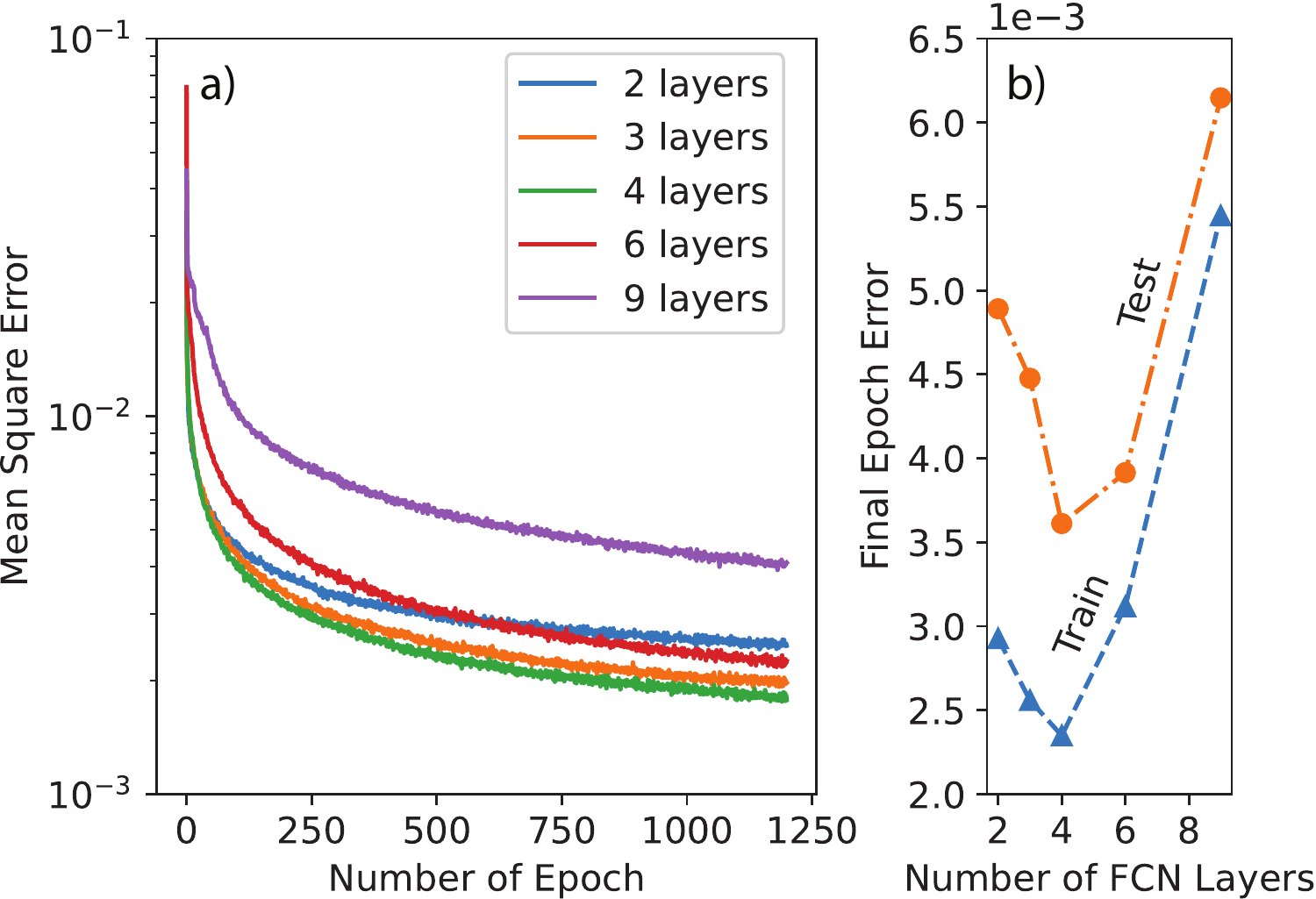}
    \caption{a) Training error curves for differing numbers of fully connected layers in the neural network. b) Training error versus test error, which is measured in logarithm (base 10) of the average mean square error per batch after 1200 epochs of training.}
    \label{fig:testing_error}
\end{figure}

We first consider networks trained on the full dataset, with varying numbers of fully connected (FCN) layers, as shown in Fig. \ref{fig:testing_error}a.
The rapid decay of each learning curve suggests that the most important optimization occurs after only a hundred epochs, but refinement of the network does continue even after a thousand epochs.
The training error with 6 FCN layers is larger than the training error with 2 FCN layers after 100 epochs' training, but the 6 FCN network outperforms the 2 FCN network after 600 epochs.
This means we can sacrifice accuracy at long training times for accuracy at short training times by decreasing the number of layers, which is a valuable feature for larger datasets.
We also compared the final epoch's test error and training error among the different number of FCN layers after 1200 epochs in Fig. \ref{fig:testing_error}b.
For this fixed training time, we find a uniform minimum in both the training and testing error at 4 FCN layers.
Therefore, in the following sections, the networks shown will always consist of 4 FCN layers.
It is also not surprising to see that the test error is always greater than the training error, as the test error consists of bilayer materials unknown to the network and some over-fitting to the training set is expected.

\begin{figure}
    \centering
    \includegraphics[width=\linewidth]{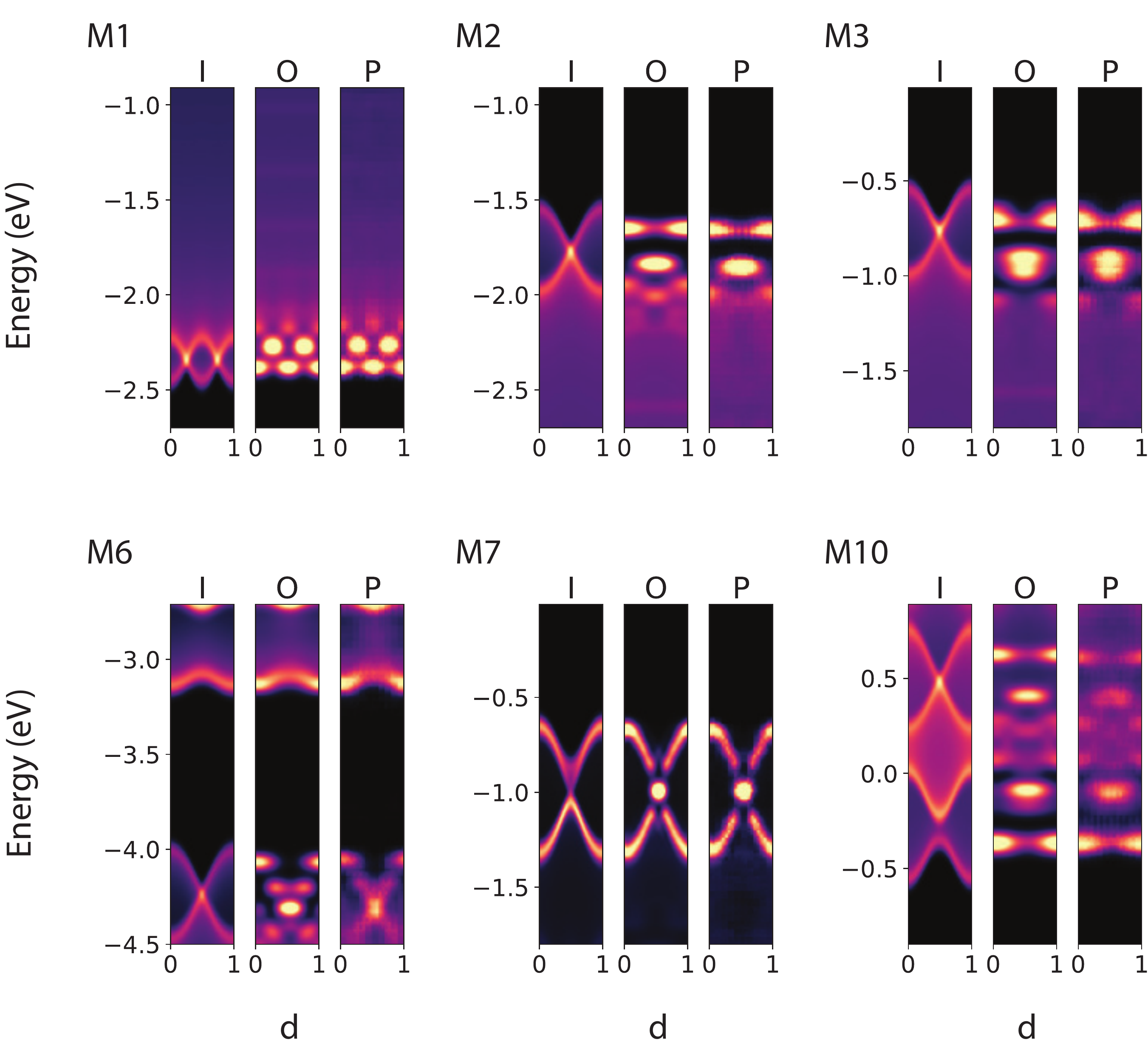}
    \caption{Comparison of configuration-dependent LDOS maps calculated from aligned bilayer Hamiltonians (input, I), moir\'e Hamiltonians (output, O), and  predictions from our CNN (P) which are generated from the supplied input after training. Six representative cases are shown here, for bilayers of material types M1, M2, M3, M6, M7, and M10.}
    \label{fig:prediction}
\end{figure}

Although the learning curves imply some success in the prediction of SD-LDOS in moir\'e bilayers, direct comparison of the predictions is more illuminating. Figure \ref{fig:prediction} presents the performance of a model trained on the full dataset for six different material samples in the test set. 
The predictions (P) and the tight-binding calculation (Output, O) for the sampled M1, M3 and M7 are nearly indistinguishable.
For the sample of M2, the primary ``bright-spot'' singularity in the predicted SD-LDOS is shifted down in energy compared to the true result, and a more moderate singularity was missed just below that.
The most complicated spectra, the sample of M10, is surprisingly reproduced quite well, with only the relative intensity of the different features slightly incorrect. 
The prediction of the M6 sample is not accurate near $-4$ eV, but this is not completely unexpected.
Because this feature is near the border of the energy window, and because a large region of this window consists of a band-gap (zero density), the network may not be gaining enough information from the input (I) to correctly predict the moir\'e structure near the window boundaries.

\begin{figure}
    \centering
    \includegraphics[width=\linewidth]{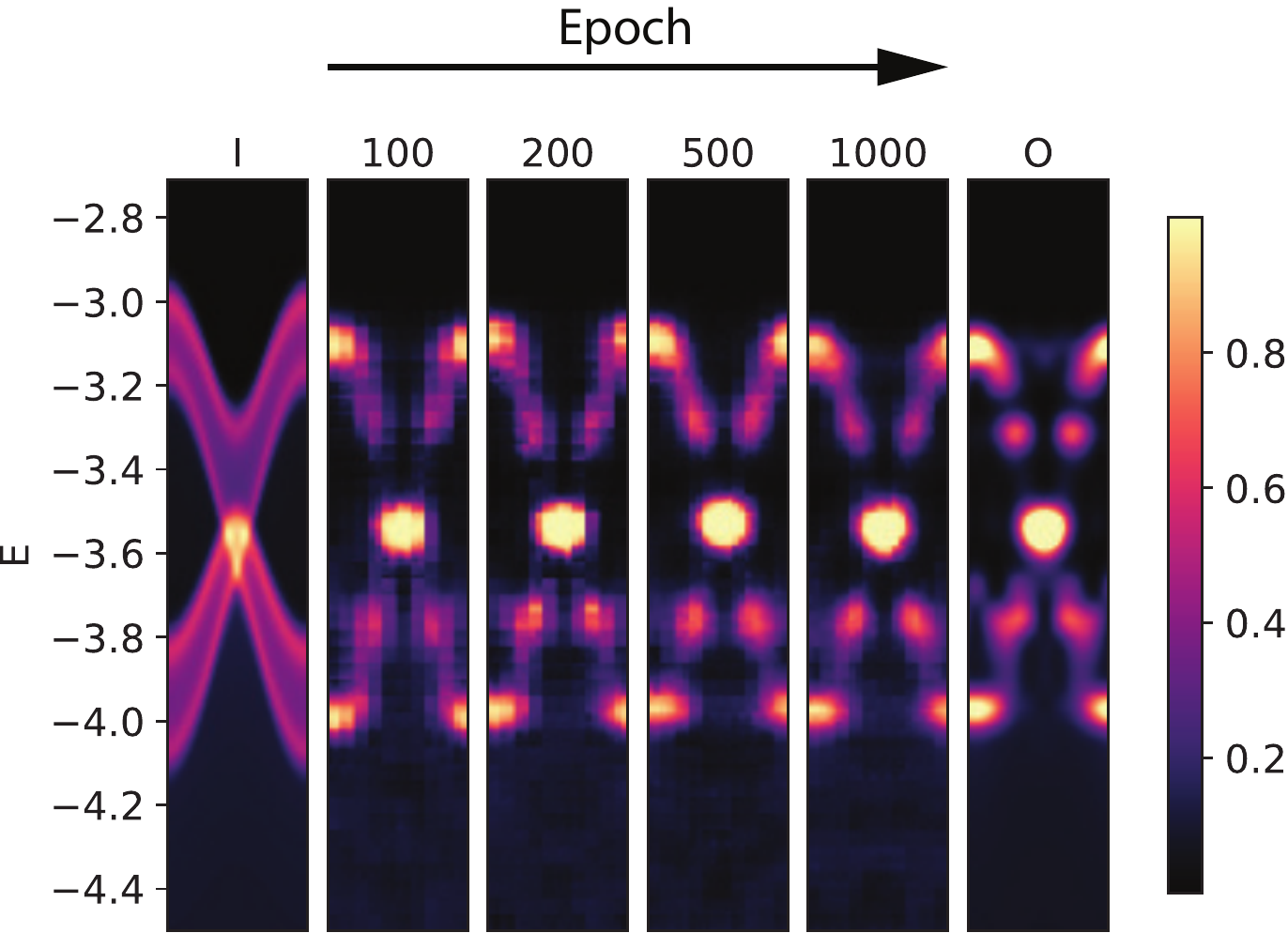}
    \caption{Evolution of the CNN prediction quality with increasing training time on an Hamiltonian sampled from M10. Input (I) is the LDOS calculated from the aligned bilayer Hamiltonian and Output (O) is the LDOS calculated from the moir\'e bilayer Hamiltonian. The panels in the middle show the predicted LDOS from neural network after corresponding 100, 200, 500, and 1000 epochs.}
    \label{fig:evolution}
\end{figure}

In Fig. \ref{fig:evolution}, we show the evolution of the network prediction during the training for an M10 sample in the testing set.
Even with just 100 epochs of training, the prediction is already good enough to easily identify the pattern of the LDOS, which corresponds with the observed fast decay of the training error in Fig. \ref{fig:testing_error}a. However, comparison between LDOS prediction after 100 epochs and 1000 epochs indicates that during the long-tail of the training, the network begins to capture the finer structure of the SD-LDOS and generate a smoother image.

\begin{figure}
    \centering
    \includegraphics[width=\linewidth]{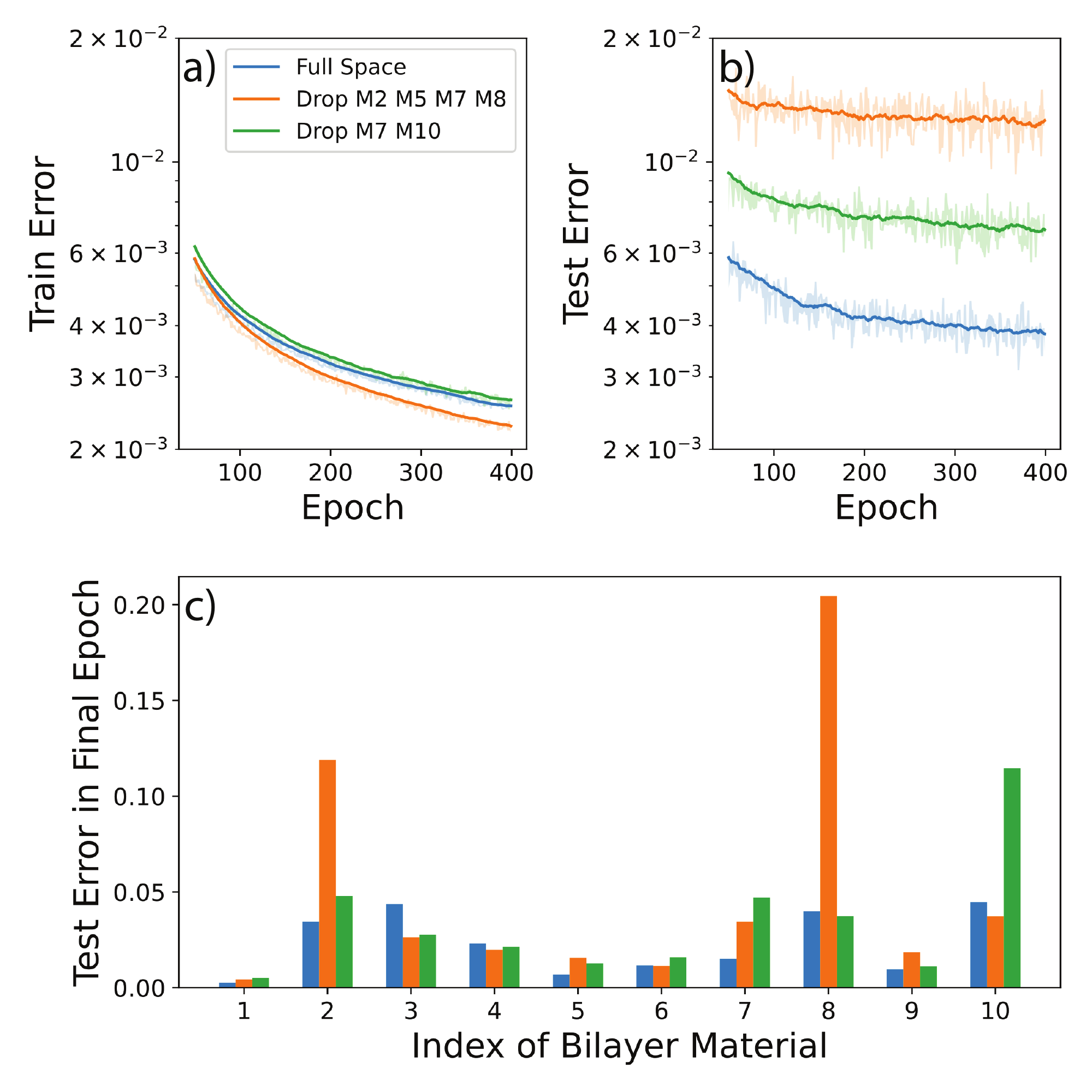}
    \caption{ a) Training error curve and b) Test error curve during the training process from 100 epochs to 400 epochs, given in mean square error (MSE) per batch. The full model as well as the two reduced models are shown in different colors. The solid lines give the error averaged over 20 epochs whereas the lighter colors give maximum and minimum error within the last three epochs. c) Comparison of the test errors for the F, R1, and R2 models, binned by the type of material, after 1200 epochs of training.}
    \label{fig:compare_re}
\end{figure}

We now turn to networks trained on the reduced datasets R1 and R2. In Fig. \ref{fig:compare_re}a, the  training errors among the three datasets are very similar.
However the testing error in Fig. \ref{fig:prediction}b, which is calculated over all 10 types of bilayer material (including those not in the R1 and R2 training sets), shows that removing materials from the training set naturally makes the performance on the full test set worse, and is roughly proportional the number of materials removed. 
We summarize the the test error after the final epoch for each of the three networks in Fig. \ref{fig:prediction}c. 
For the material types M2, M5, M7, M8, and M10, which are dropped at least once, it can be seen that a reduced model trained without samples of that material type can cause up to four times larger total MSE for that material.

\begin{figure}
    \centering
    \includegraphics[width=\linewidth]{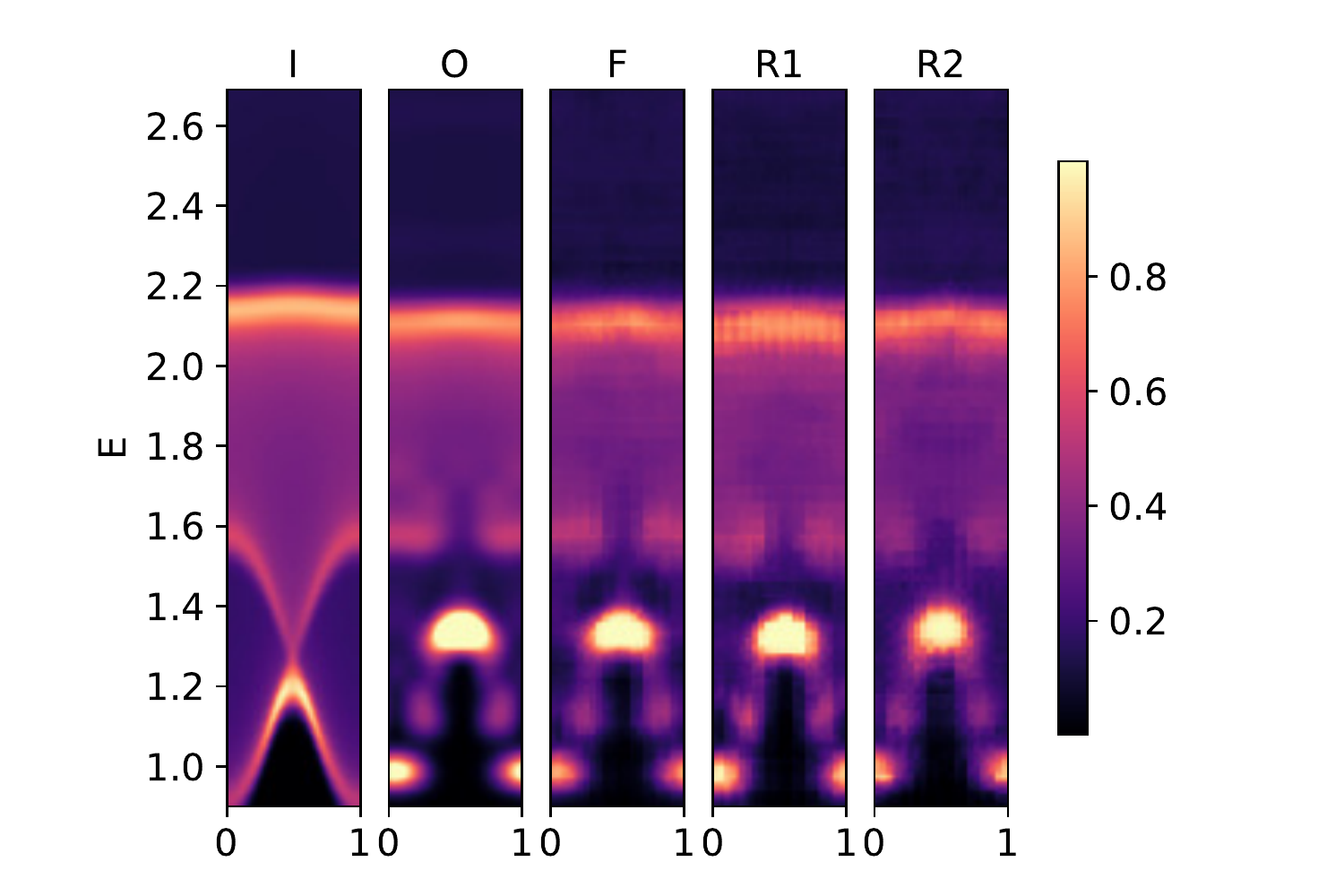}
    \caption{Comparison of the full training model and reduced training models on a Hamiltonian sampled from M2. Input (I) is LDOS calculated from the aligned bilayer Hamiltonian and Output (O) is the SD-LDOS calculated in the moir\'e bilayer Hamiltonian, both via exact diagonalization. The last three panels shows the predicted moir\'e SD-LDOS from CNN models based on different training sets: F is the full training model which includes all materials, R1 the model where M7 and M10 are dropped in the training set, and R2 the model when M2, M5, M7 and M8 are dropped.}
    \label{fig:reduce}
\end{figure}

However, the increase in the MSE does not tell the full story of this transfer learning test. 
Let's investigate how the different networks performed on a sample of the M2 material in the test set, as shown in Fig. \ref{fig:reduce}.
As M2 was not in R2's training set, but was in the training for R1 and the full (F) set, one might expect that the result from the R2 network would be signifcantly worse than the other two. However, all three networks perform quite well on this M2 sample.
So although the average MSE of M2 by the R2 network is higher, transfer learning appears to be occurring in this example.
     
\begin{figure}
    \centering
    \includegraphics[width=\linewidth]{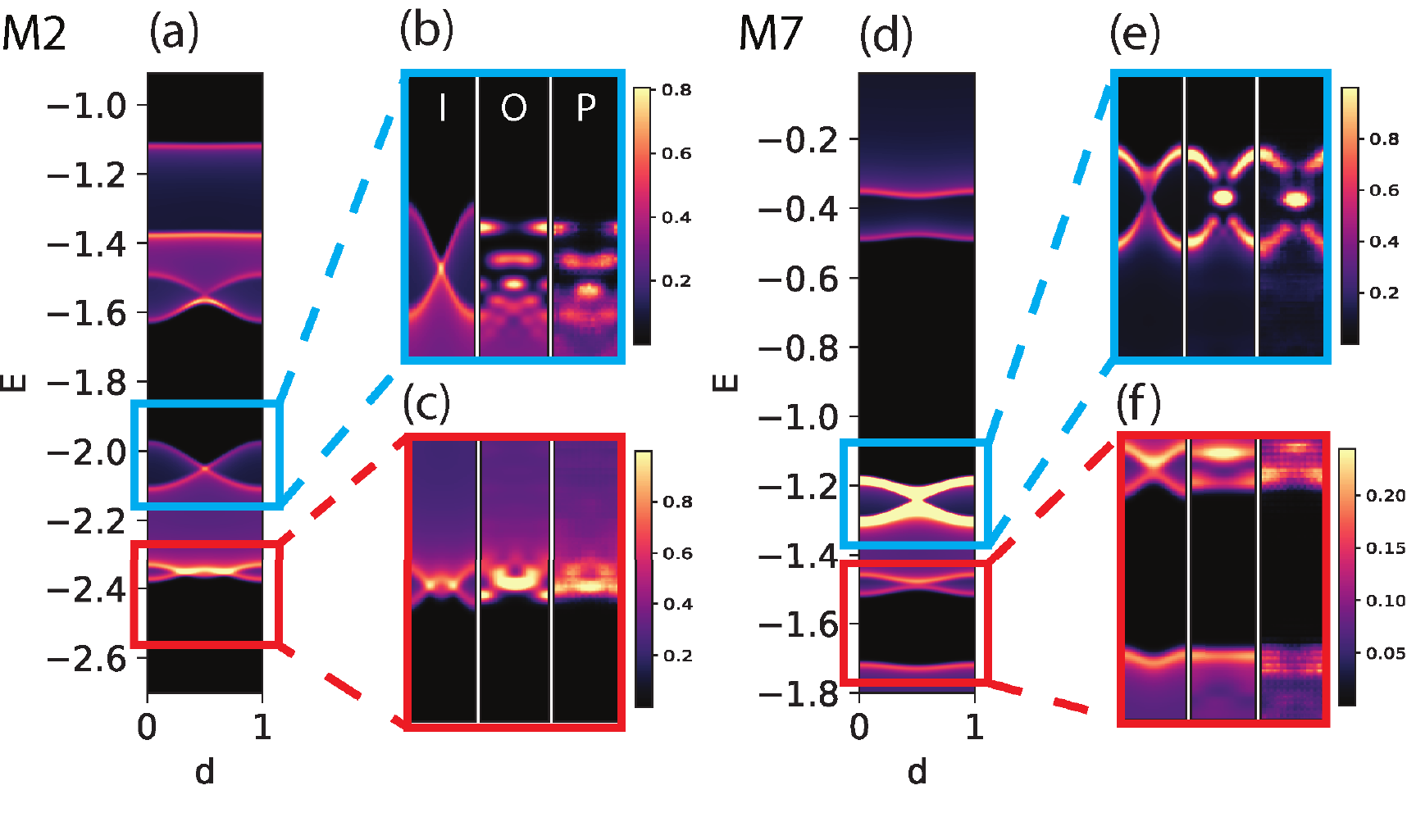}
    \caption{Predictions by the reduced model R2 (dropping M2, M5, M7, and M8) of one sample of M2 and one of M7. (a) and (d) are the full spectrum stacking dependent local density of states (SD-LDOS) of the aligned bilayer material from M2 and M7, respectively. (b) and (c) show the SD-LDOS input (I), output (O), and network prediction (P) for two selected windows of the M2 spectrum, with (e) and (f) the same but for the M7 spectrum.}
    \label{fig:reduced_R2}
\end{figure}

The predictions of the SD-LDOS by the R2 network on two samples of materials not in its training set are presented in Fig. \ref{fig:reduced_R2}. In principle, dropping almost half of the full dataset should lead to a poor network prediction, as implied from higher MSE error in Fig. \ref{fig:compare_re}. This is evident to some extent in Fig. \ref{fig:reduced_R2}c, which shows a poor prediction by the network on this unseen material.
Indeed, many examples can be found from the dropped material types which show large errors, arising from an incorrect energy shift or LDOS height of the singularity, and in many cases a complete failure to predict a similar SD-LDOS pattern. However, there still exists some evidence of effective transfer learning in this task. 
Fig. \ref{fig:reduced_R2}b,e show examples of good neural net predictions on the unseen material types.
There are also numerous predictions where the singularity is not perfectly reconstructed, for example with a slight energy shift as in Fig. \ref{fig:reduced_R2}f, but is still good enough for a screening task.
Our reduced model can predict LDOS well on some of the unseen data because similar band structures within a truncated energy window exist across the training data. Successful learning of the local electronic structure is a clear suggestion of effective transfer learning, and the larger MSE in M7 and M10 of model R1 is simply a sign that band structures similar to M7 and M10 are not present in the rest of the data, which was indeed the guiding motivation in defining that reduced model.

\begin{figure}
    \centering
    \includegraphics[width=\linewidth]{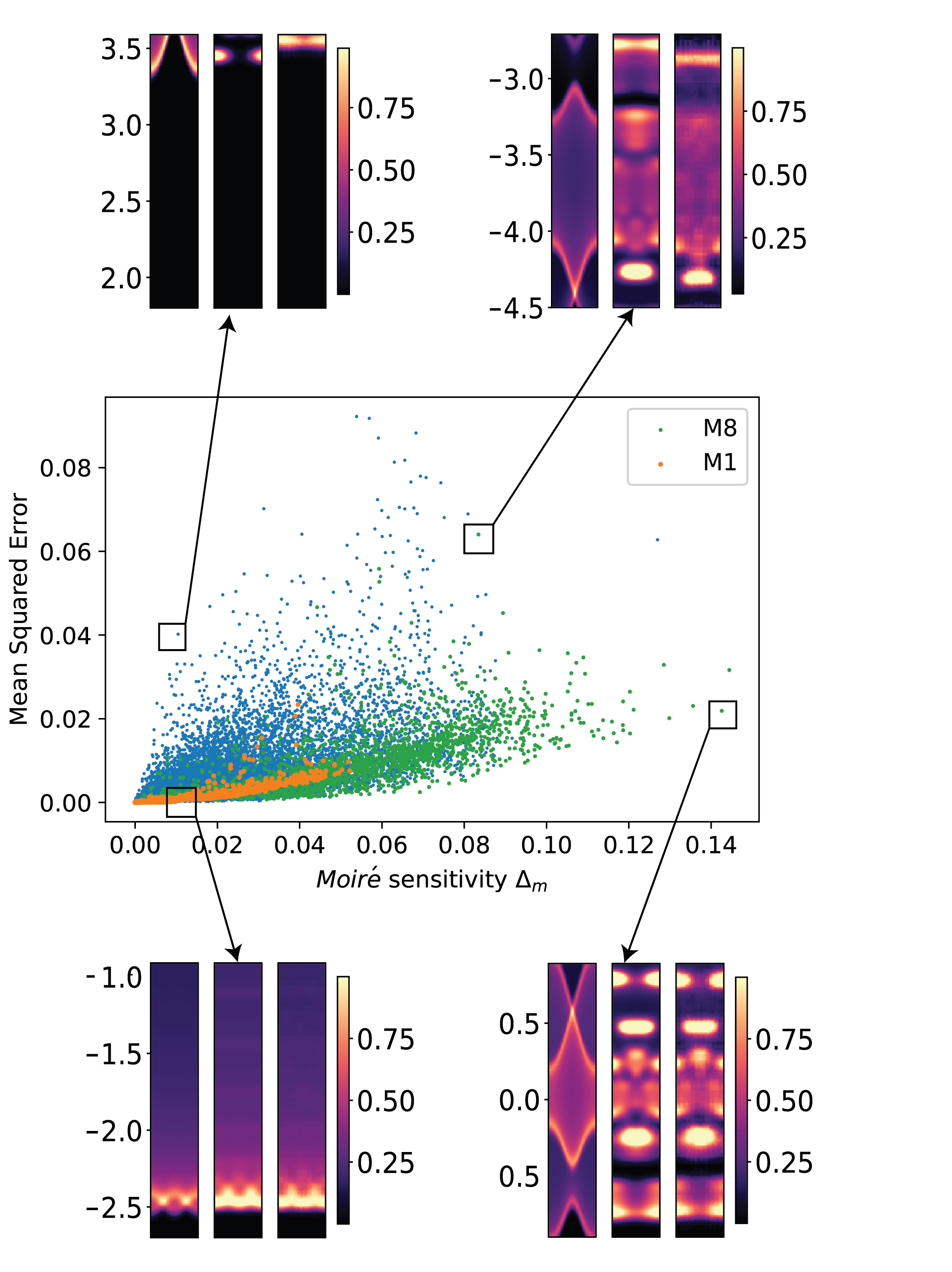}
    \caption{CNN prediction error as a function of each window's electronic structure sensitivity to moir\'e structure, $\Delta_m$, trained on the full data set (F). Each point corresponds to a specific material and spectral window sample in the test set. The input (I), output (O), and CNN predicted (P) SD-LDOS for four selected points are shown above and below the main plot, with the three panels of each inset in that respective order.}
    \label{fig:senstivity_error}
\end{figure}

To better understand whether the learning from the full data set is effective and why the R1 and R2 networks have reduced performance on average, but on many specific material samples still perform quite well, we will need to investigate the distribution of the MSE contribution from the test set.
In Fig. \ref{fig:senstivity_error}, we investigate how the MSE and the sensitivity of the moir\'e electronic structure are related ($\Delta$ and $\Delta_m$ respectively, as defined in Eq. \ref{eq:error}).
As a reminder, large moir\'e sensitivity $\Delta_m$ corresponds to materials which have large differences in their aligned and moir\'e electronic structures.
Even for samples with substantial moir\'e sensitivity ($\Delta_m > 0.08$), the prediction MSE could still be small.
Furthermore, we find that the samples of M8 (blue) tend to be the hardest to predict and have large moir\'e sensitivity.
In contrast, the samples of M1 (orange) are the easiest to predict and tend to have the smallest moir\'e sensitivity. This helps explain why M8 shows the highest amount of transfer learning error (in the orange MSE of Fig. \ref{fig:compare_re}) while M1 shows the smallest learning error across all models.

There are four specific material samples emphasized in the insets of Fig. \ref{fig:senstivity_error}, comparing the true SD-LDOS to the network prediction. The upper panels represent cases where the error is relatively large compared to $\Delta_m$. For both these cases, the error is mostly due to an energy offset between the true and predicted LDOS singularity features. However, ignoring this relative shift the qualitative prediction of the CNN is still quite good even though the MSE loss function considers both of these predictions as poor.
Within this context, we expect the increased MSE on the removed materials in R1 and R2 is also caused by the networks inability to properly predict the relative shift in singular SD-LDOS features.
The bottom left inset of Fig. \ref{fig:senstivity_error} is an example where both the error and moir\'e sensitivity are small. Here the small sensitivity is due to a large amount of shift-independent LDOS in the aligned system, but the prediction of the moir\'e features (near -2.5 eV) is still good.
The bottom right panel shows the case where $\Delta_m$ is large, but the error is small, and is among the most impressive predictions from this testing set. The patterns of moir\'e SD-LDOS here are very complex, but the neural network predicts it accurately. The scattering plot also suggests that even when the MSE error is large, the LDOS prediction could be good enough as a screening tool for correlated materials.


\section{Conclusion}
\label{sec:conclusion}

We have trained a convolution neural network to predict moir\'e electronic structure, which may allow for the identification of good candidates for correlated phases from 2D material databases. Our CNN uses an encoder-decoder structure, with a fully connected network at the center. To test the viablity of such an approach, we have focused on the simplified problem of learning 1D moir\'e electronic structures at a fixed moir\'e mismatch of $\theta = 0.1$, and found that the network can predict the SD-LDOS of 1D materials.
Note that much of the useful information that is obtained from exact treatment of the moir\'e electronic structure problem, such as explicit band structures and symmetry representations of those bands, does not appear obtainable with this approach. We see our approach as a powerful addition to the scientific toolbox, but emphasize that it is by no means a complete replacement for existing approaches.

Ten classes of artificial bilayer structures were designed and thousands of individual materials were generated from randomization of tight-binding parameters. When trained on the full data set, the neural network makes excellent predictions of moir\'e electronic structure. In most cases, we find accurate predictions of singularities in the SD-LDOS, which indicate the energy and localization pattern of moir\'e flat bands. The shape and scale of such singularities is  inaccurate in some instances, possibly caused by failures in data preprocessing or unfortunate energy window cutoffs. When training on the full dataset, the learning curves show that a more complex network may give better predictions but would require a much longer training time.

We also investigate the network's ability to make predictions on materials outside of its training set by using two different reduced data sets which drop specific classes of the 1D moir\'e materials. By comparing these models' results with the model trained on the full data set, we find that the neural network predictions for unfamiliar materials are still reliable. This supports our hypothesis that there exists a universal twist operator, and confirms that moir\'e electronic structure can be captured by the local structure of the aligned SD-LDOS.
The overall performance of the network is best understood by comparison of the prediction error $\Delta$ to the moir\'e sensitivity $\Delta_m$.
We find that the majority of our validation set has $\Delta < \Delta_m/5$, with the network performing performing significantly better on some materials over others.

Although this work has shown the effectiveness of applying neural networks to learn moir\'e electronic structure, the mathematical theory behind the framework is still not fully explored. A clearer mathematical framework for describing operator learning of moir\'e electronic structure would help clarify and improve any machine learning application on this topic.
To extend this methodology to also take a given twist angle, $\theta$, as an additional input, the most straightforward approach would be to train multiple networks for each $\theta$ in a set of angles, and then interpolate results between them.
A more sophisticated, and likely better performing approach, is to insert the parameter $\theta$ as a special parameter to the network, perhaps as a linear operation on each layer of the network instead of just in the initial input layer.
This would allow the entire network to have a smooth dependence on the desired twist-angle, including the filtering functions and the fully connected layers, allowing for transfer learning between data at different angles.

Moreover, since the neural network framework has been able to predict the 1D moir\'e electronic structure, it is natural to extend the framework to 2D moir\'e material. Although there are no conceptual difficulties in this generalization, the elements of the datasets will be three dimensional (volumetric data) instead of two dimensional (images), likely leading to significantly longer training times. The performance of neural networks can be tested on an artificial library of materials in the 2D case, but now comparison to first principles calculations and experimental data become possible. Another key ingredient in the electronic structure of 2D moir\'e material is relaxation of the atomic structures. It is straightforward to incorporate relaxation into the neural network based method by replacing the unrelaxed twisted SD-LDOS with data from relaxed calculations, but its effect on the training speed and prediction accuracy are uncertain.

\begin{acknowledgments}
The calculations in this work were performed using computational resources and services at the Center for Computation and Visualization (CCV) at Brown University and the Minnesota Supercomputing Institute (MSI) at the University of Minnesota.
DL was supported in part by NSF DMREF Award No. 1922165.
ML was supported in part by NSF DMREF Award No. 1922165 and Simons Targeted Grant Award No. 896630.
SC was supported by the National Science Foundation under grant No. OIA-1921199.
\end{acknowledgments}

\appendix
\counterwithin{figure}{section}
\counterwithin{table}{section}

\section{1D Hamiltonians and band calculations}
\label{app:hams}

Each 1D material is defined by three objects: the hopping terms between orbitals within the same unit cell ($H_0$), hopping terms to orbitals in the unit cell to the right ($H_x$), and the orbital positions ($A = \{\vec{r}_1, \vec{r}_2, \dots \}$, with the $i$'th column of $A$ giving the $(x,y)$ location of atom $i$).
The electronic hopping terms to orbitals in the left-neighboring unit cell are given by $H_{-x} = H_x^\dagger$.
For example, if there are three orbitals in the unit cell of the 1D material, then $H_0$ and $H_+$ are $3 \times 3$ matrices. If we construct a 1D supercell consisting of ten units, then the full Hamiltonian for that layer will be a $30 \times 30$ matrix that depends on the crystal momentum $\vec{k}$.
We assume each monolayer has an unstrained lattice parameter of $\alpha = 1$, and that each bilayer has an interlayer distance $z = 1$ between the layers. Every orbital in a given layer has the same $z$-coordinate (e.g. there is no vertical difference between the atomic positions).

\begin{table*}[]
    \centering
    \begin{tabular}{|c||c|c|c|c|c|c|c|c|c|c|}
        \hline
            M$i$ & M1 & M2 & M3 & M4 & M5 & M6 & M8 & M9\\ \hline
        $t^i_1$ & $1.25 \pm 0.25$ & $2 \pm 1$ & $-2 \pm 1$ & $-2 \pm 1$ & $2 \pm 1$ & $2 \pm 1$ & $-1.75 \pm 0.75$ & $2.5 \pm 1$ \\ \hline
        $t^i_2$ & 0 & 0 & 0 & 0 & $1.5 \pm 0.5$ & $1.5 \pm 0.5$ & 0 & $1.5 \pm 1$ \\ \hline
        $o^i_1$ & 0 & $-1.5 \pm 0.5$ & $0 \pm 1$ & $-1.5 \pm 0.5$ & $-1.5 \pm 0.5$ & $0.5 \pm 1.5$ & $-3.5 \pm 1$ & $2.5 \pm 1.5$ \\ \hline
        $o^i_2$ & 0 & $1.5 \pm 0.5$ & $2.5 \pm 1.5$ & $0.5 \pm 0.5$ & $0.5 \pm 0.5$ & $-4 \pm 2$ & $-2 \pm 1.5$ & $-4 \pm 2$ \\ \hline
    \end{tabular}
    \caption{Tight-binding parameters for all materials except M7 and M10. All units in eV.}
    \label{tab:simple_ham_params}
\end{table*}

\begin{figure}
    \centering
    \includegraphics[width=\linewidth]{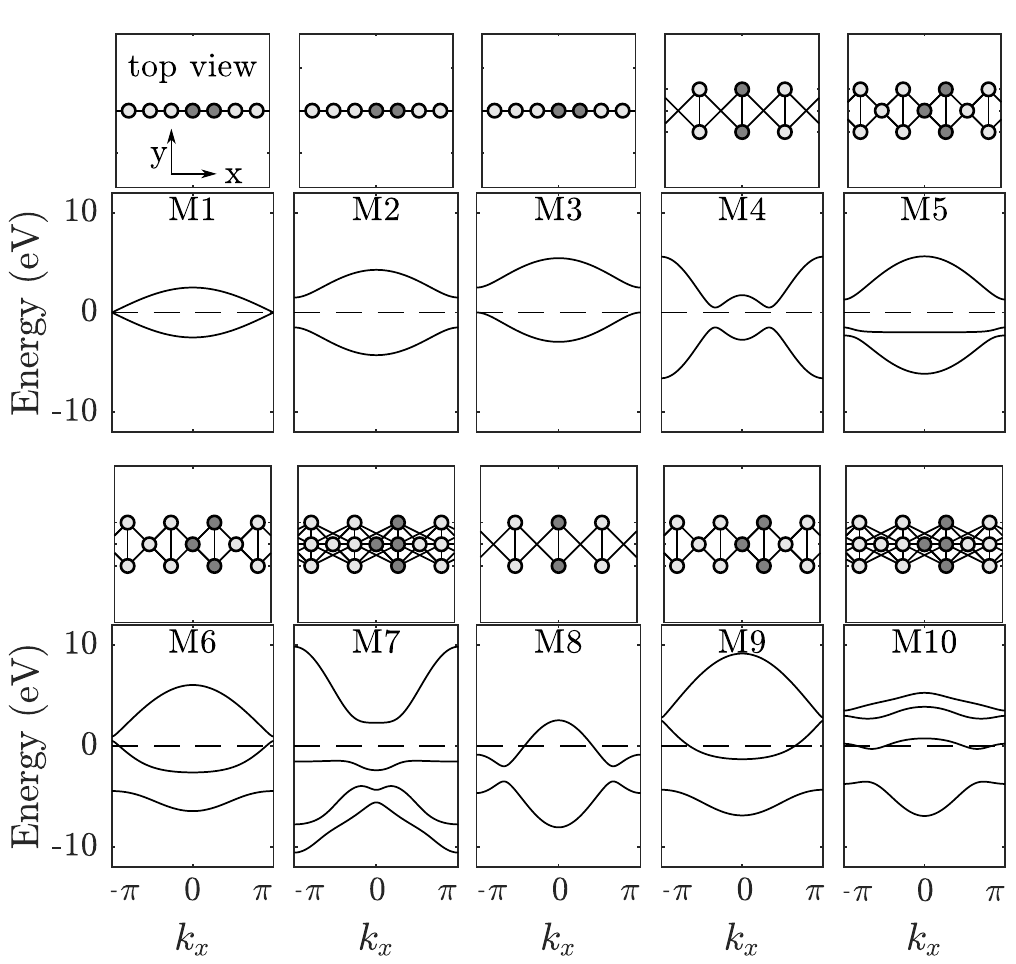}
    \caption{Monolayer atomic geometry and band structures for the ten 1D materials used in this work. The geometry is displayed from a top-down view, with the atoms in the primitive unit-cell colored in dark-grey and with each line between atoms representing an electronic hopping in the tight-binding model.}
    \label{fig:monolayer_bands}
\end{figure}

We calculate electronic structure efficiently using Bloch's theorem for a given 1D crystal momentum $k_x$. We include the phase by modifying the hopping element from orbital $j$ to orbital $i$ according to
\begin{equation}
    t_{ij} \to t_{ij} e^{i k_x r^{ij}_x}
\end{equation}
where $r^{ij}_x$ is the $x$-distance from orbital $j$ to orbital $i$.
If $t_{ij}$ corresponds to a coupling which traverses the supercell's periodic boundary condition, then it is understood that $\vec{r}_{ij}$ is taken as if one of the orbitals has been moved to its periodic partner in the neighboring supercell.
Interlayer couplings, between orbitals $i$ and $j$ of neighboring chains, are given by Eq. \ref{eq:interlayer_t} of the main text, with $\vec{r} = \vec{r}^{ij}$.

We now provide the details of the Hamiltonians for each of the ten material-types.
The material index is given as a superscript of each matrix or parameter, for example $H_0^{1}$ and $t_1^{1}$ correspond to the onsite Hamiltonian and first tunneling parameter of M1, respectively.
The tunneling ($t$) and onsite energies ($o$) for all materials are given in Tab. \ref{tab:simple_ham_params} and \ref{tab:complex_ham_params}, with parameters given as $a \pm b$ to indicate that the parameter is randomly sampled from a uniform distribution in the range $[a-b, a+b]$.
The atomic geometry and band structure for each monolayer is shown in Fig. \ref{fig:monolayer_bands}.

For M1, M2, and M3, the Hamiltonians are given by
\begin{equation}
        H^{i}_0 = 
        \begin{pmatrix}
            o^i_1 & t^i_1 \\
            t^i_1 & o^i_2 \\
        \end{pmatrix},
        H^{i}_x =
            \begin{pmatrix}
            0 & t^i_1 \\
            0 & 0 \\
        \end{pmatrix},
        A^{i} =
            \begin{pmatrix}
            0 & 0.5 \\
            0 & 0 \\
        \end{pmatrix}.
\end{equation}

The M4 and M8 Hamiltonians are given by:

\begin{equation}
        H^{i}_0 = 
        \begin{pmatrix}
            o^i_1 & t^i_1 \\
            t^i_1 & o^i_2 \\
        \end{pmatrix},
        H^{i}_x =
            \begin{pmatrix}
            0 & t^i_1 \\
            t^i_1 & 0 \\
        \end{pmatrix},
        A^{i} =
            \begin{pmatrix}
            0 & 0 \\
            -0.5 & 0.5 \\
        \end{pmatrix}.
\end{equation}

The M5, M6, and M9 Hamiltonians are given by:
\begin{equation}
        H^{i}_0 = 
        \begin{pmatrix}
            o^i_1 & t^i_1 & t^i_1 \\
            t^i_1 & o^i_2 & t^i_2 \\
            t^i_1 & t^i_2 & o^i_1
        \end{pmatrix},
        H^{i}_x =
            \begin{pmatrix}
            0 & t^i_1 & t^i_1 \\
            0 & 0 & 0 \\
            0 & 0 & 0
        \end{pmatrix},
        A^{i} =
            \begin{pmatrix}
            0 & 0.5 &  0.5 \\
            0 & 0.5 & -0.5\\
        \end{pmatrix}.
\end{equation}

\begin{table}[]
    \centering
    \begin{tabular}{|c||c|c|}
        \hline
            M$i$ & M7 & M10\\ \hline
        $t^i_1$ & $2.5 \pm 1.5$ & $-1 \pm 0.5$ \\ \hline
        $t^i_2$ & $-1.5 \pm 0.5$ & $1 \pm 0.5$ \\ \hline
        $t^i_3$ & $1.5 \pm 0.5$ & $-1.5 \pm 0.5$ \\ \hline
        $o^i_1$ & $-4.5 \pm 1.5$ & $3 \pm 1$ \\ \hline
        $o^i_2$ & $-3 \pm 1$ & 0 \\ \hline
        $o^i_3$ & $-4.5 \pm 1.5$ & $3 \pm 2$ \\ \hline
        $o^i_4$ & $2 \pm 2$ & $-3 \pm 2$ \\ \hline
    \end{tabular}
    \caption{Tight-binding parameters for M7 and M10.}
    \label{tab:complex_ham_params}
\end{table}

Finally, M7 and M10 are given by

\begin{equation}
    \begin{split}
        H^{i}_0 &= 
        \begin{pmatrix}
            o^i_1 & t^i_2 & t^i_1 & t^i_1 \\
            t^i_2 & o^i_2 & t^i_2 & t^i_2 \\
            t^i_1 & t^i_2 & o^i_3 & t^i_3 \\
            t^i_1 & t^i_2 & t^i_3 & o^i_4
        \end{pmatrix}, \\
        H^{i}_x &=
            \sigma^i \begin{pmatrix}
            0 & t^i_2 & t^i_1 & t^i_1 \\
            0 & 0 & t^i_2 & t^i_2 \\
            0 & t^i_2 & 0 & 0 \\
            0 & t^i_2 & 0 & 0
        \end{pmatrix}, \\
        A^{i} &=
            \begin{pmatrix}
            0 & 0.5 &  0.5 & 0.5 \\
            0 & 0 & -0.5 & 0.5 \\
        \end{pmatrix},
    \end{split}
\end{equation}

with $\sigma^i = -1$ for M7, and $\sigma^i = 1$ for M10.

\bibliography{moire}

\end{document}